\documentclass[final]{svjour2}
\usepackage{graphicx}
\usepackage{rotating}
\usepackage{amssymb}
\usepackage{mathptmx}
\usepackage[numbers]{natbib}
\makeatletter
\journalname{Journal of Low Temperature Physics}
\bibpunct{}{}{,}{s}{}{,}

\begin{document}

\newcommand{\hdblarrow}{H\makebox[0.9ex][l]{$\downdownarrows$}-}
\title{Modal Decomposition in Goalpost Micro/nano Electro-mechanical Devices}

\author{E. Collin \and M. Defoort \and K.J. Lulla  \and J. Guidi \and S. Dufresnes  \and H. Godfrin}
% \and C. Blanc \and O. Bourgeois
\institute{Institut N\'{e}el, CNRS et Universit\'{e} Joseph Fourier,\\ BP 166, 38042 Grenoble Cedex 9, France\\
\email{eddy.collin@grenoble.cnrs.fr}}

\date{19.06.2012}
\maketitle


\begin{abstract}

We have studied the first three symmetric out-of-plane flexural resonance modes of a goalpost silicon micro-mechanical device.
Measurements have been performed at 4.2$~$K in vacuum, demonstrating high $Q$s and good linear properties. 
Numerical simulations have been realized to fit the resonance frequencies and produce the mode shapes. 
These mode shapes are complex, since they involve distortions of two coupled orthogonal bars. 
Nonetheless, analytic expressions have been developed to reproduce these numerical results, with no free parameters.
Owing to their generality they are extremely helpful, in particular to identify the parameters
which may  limit the performances of the device.
The overall agreement is very good, and has been verified on our nano-mechanical version of the device.

\end{abstract}

\section{Introduction}

Vibrating objects are a common tool used in low temperature physics in order to probe the intriguing properties of quantum fluids.
One of the first devices employed has been the so-called vibrating wire, which has enabled the direct measurement of the coldest temperatures in superfluid He$^3$ ever reported \cite{lancaster}. Other devices have been used like oscillating spheres \cite{schoepe}, and today  an increasing number of physicists are taking advantage of the very practical quartz tuning fork technique \cite{forks1}.  

The Grenoble group has been following another path: the idea being taking advantage of the versatility of microfabrication techniques \cite{Physica_2000}.
We have been developing and testing "goalpost" shaped silicon devices, in both a MEMS$^($\footnote{MEMS: micro-electro-mechanical-systems.}$^)$ and NEMS$^($\footnote{NEMS: nano-electro-mechanical-systems.}$^)$ variant having respectively (transverse) dimensions ranging from a few microns to a hundred nanometers \cite{JLTP_2008,QFS_2010}. They both display extremely promising mechanical properties, with for the first out-of-plane flexural mode high quality factors and frequencies lying within typically $1~$kHz up to $10~$MHz.
 
All of these structures also have higher flexural modes.  These modes have rather complex shapes, but present a very simple way to probe the fluid at substantially different frequencies. They can also be thought of as an opportunity to set up new measuring schemes, like driving one mode (at one frequency) and measuring another one (having a very different frequency), mimicking the setup used for so-called mode-coupling mechanical experiments \cite{kunal}.

In the present paper we report on experiments performed on a MEMS (Al on silicon) goalpost device, in vacuum at 4.2$~$K. During this work we also measured a NEMS device (made of the same materials), reaching the same conclusions. The excitation and detection is performed with the well known magnetomotive scheme. 
Details, including the fabrication processes can be found in Refs. \cite{JLTP_2008,QFS_2010}. 
We have performed numerical simulations enabling to understand the flexural mode shapes, and to predict the measured resonance frequencies of the 3 first symmetric modes.
An analytic model is given reproducing these numerical results, with no free parameters. Moreover, the quality factors of these modes appear to be independent of the mode number, being all of the order of $0.2 \times 10^6$ at Helium temperatures.

\begin{figure}
\begin{center}
\includegraphics[width=0.9\linewidth,keepaspectratio]{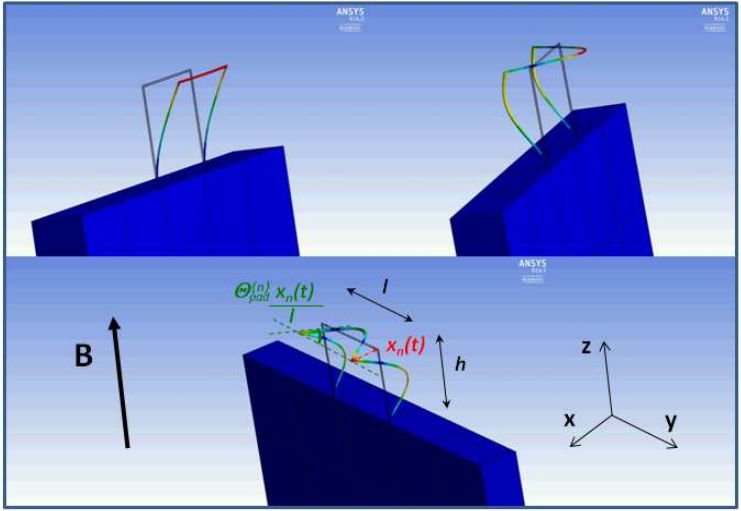}
\end{center}
\caption{(Color online) First out-of-plane symmetric mode shapes. From top left clockwise: first mode, third mode, and fifth mode (with details about the modeling shown). Dimensions are (same as actual device): length pad $l=1.2~$mm, length foot $h=1~$mm, width pad $w'=15~\mu$m, width foot $w=15~\mu$m, silicon thickness $e=13.5~\mu$m, and metal thickness $e_M=100~$nm.}
\label{fig1}
\end{figure}

\section{Theoretical mode description}

In Fig. \ref{fig1} we present the mode shapes computed from numerical finite element analysis (plus parameters used in the theoretical modeling). We used the ANSYS \cite{ansys} platform.
The silicon has been defined as orthotropic with elastic parameters taken from the literature \cite{youngs}. The aluminum layer is isotropic, with Young's modulus $E_m=80~$GPa and Poisson's ratio $\nu = 0.25$. The modeling does not incorporate any in-built stresses.

From simulations we realize that the out-of-plane flexure essentially reproduces the natural modes of cantilevers (the feet, in common motion) loaded at the end by the force exerted by the distortion of the paddle. The motion of the two feet is in-phase for symmetric modes, and out-of-phase for anti-symmetric ones with respect to a plane perpendicular to the mid-point of the paddle. 

To compute analytically the mode shapes, let us first define the distortion of the paddle $f_{pad}^{\{n\}}(y,t)$ for (foot) mode $n\geq1$. This part of the device is essentially a doubly-clamped beam satisfying the Euler-Bernoulli equation, with boundary conditions:
\begin{eqnarray*}
\frac{\partial f_{pad}^{\{n\}} (y=0,t)}{\partial y}  & = &  \theta_{pad}^{\{n\}} \, \frac{x_n(t)}{l}, \\  
\frac{\partial f_{pad}^{\{n\}} (y=l,t)}{\partial y}  & = & -(-1)^{n+1} \theta_{pad}^{\{n\}} \, \frac{x_n(t)}{l} ,
\end{eqnarray*}
the mode number $n$ being odd for symmetric, and even for anti-symmetric modal shapes. $x_n(t)$ is the displacement of the end part of one foot
(let it be the right one), while $(-1)^{n+1} \, x_n(t)$ corresponds to the other one. One foot flexure (left or right) is thus expressed by $f_{foot}^{\{n\}}(z,t)=\pm  x_n(t) \, \Psi_{foot}^{\{n\}}(z)$. This imposes two other boundary conditions:
%\vspace*{-3mm}
\begin{eqnarray*}
f_{pad}^{\{n\}} (y=0,t)  & = &  x_n(t), \\  
f_{pad}^{\{n\}} (y=l,t)  & = & (-1)^{n+1} x_n(t) .
\end{eqnarray*}
Forcing the symmetry or anti-symmetry of the solution, we obtain for the former case:
\begin{eqnarray}
f_{pad}^{\{n\}}(y,t) & = & x_n(t) \, \Psi_{pad}^{\{n\}}(y) , \nonumber \\
\!\!\!\!\!\!\!\!\!\!\!\!\!\!\!\!\!\!\!\!\!\!\!\!\!\!\!\!\!\!\!\!\!\!\!\!\!\!\!\!\!\!\!\! \Psi_{pad}^{\{n\}}(y) &\!\!\!\!\! = & \!\!\!\!\!\!  \left[ \lambda' \sin\left( \frac{\lambda'}{2}\right) \cosh\left[ \frac{\lambda'-2 \lambda' \left(y/l\right)}{2}\right]+ 
\lambda' \sinh\left( \frac{\lambda'}{2}\right) \cos\left[ \frac{\lambda'-2 \lambda' \left(y/l\right)}{2}\right] \right.  \nonumber \\
& \!\!\!\!\!\!\!\!\!\!\!\!\!\!\!\!\!\!\!\!\!\!\!\!\!\!\!\! + &\!\!\!\!\!\!\!\!\!\!\!\!\!\!\!\!\! \left. \theta_{pad}^{\{n\}} \left( \cosh\left( \frac{\lambda'}{2}\right) \cos\left[ \frac{\lambda'-2 \lambda' \left(y/l\right)}{2}\right] -
 \cos\left( \frac{\lambda'}{2}\right) \cosh\left[ \frac{\lambda'-2 \lambda' \left(y/l\right)}{2}\right] \right) \right] \nonumber \\
& / & \left[ \lambda' \sin\left( \frac{\lambda'}{2}\right) \cosh\left( \frac{\lambda'}{2}\right) + 
\lambda' \sinh\left( \frac{\lambda'}{2}\right) \cos\left( \frac{\lambda'}{2}\right)\right], \label{padshape}
\end{eqnarray}
having defined:
\begin{eqnarray}
\lambda' & = & \left( \frac{E_z}{E_y} \right)^{1/4} \frac{l}{h} \, \lambda^{\{n\}}. \label{eqnlambda}
\end{eqnarray}
$l$ is the length of the paddle while $h$ is the length of one foot. The thickness of the structure is $e$ and $w$, $w'$ are the widths of foot and paddle respectively; $I_y=\frac{1}{12} w' e^3$ is the second moment of area of the paddle while $I_z=\frac{1}{12} w e^3$ is the one of one foot, and $E_y,E_z$ is Young's modulus along the corresponding $\vec{y},\vec{z}$ axes. $\lambda^{\{n\}}$ is the mode number defining the resonance frequency $\omega_n$ of the feet through $\omega_n =  \left(\lambda^{\left\{n\right\}} \right)^2 \, \sqrt{\frac{E_z\,I_z / h^3}{\rho \,(e w) h}}$. Note that $\lambda'$ {\it is not} resonant with the natural modes of the paddle.

The torsion equation of one foot defines the paddle's boundary angle $\theta_{pad}^{\{n\}}$:
\begin{equation}
\frac{\partial^2 \Theta (z,t)}{\partial t^2} = \frac{G_z}{\rho} \, \frac{\partial^2 \Theta (z,t)}{\partial z^2}, \label{torsionEq}
\end{equation}
with $\Theta (z,t)$ the torsion angle, $G_z$ the torsion modulus taken along the $\vec{z}$ axis, and $\rho$ the mass density.
Eq. (\ref{torsionEq}) is easily solved for a harmonic solution $\Theta_{n} (z,t)$ with boundary conditions:
%\vspace*{-5mm}
\begin{eqnarray*}
\Theta_{n} (z=0,t) & = & 0 , \\
G_z I_\theta \, \frac{\partial \Theta_{n} (z=h,t)}{\partial z} & = & M_{pad}^{\{n\}} (t) , 
\end{eqnarray*}
with $I_\theta$ the associated second moment of area, and $M_{pad}^{\{n\}}(t)$ the torsion moment exerted by the paddle, in mode $n$ of the feet. By definition $\Theta_{n} (z=h,t)=\Theta_{max}^{\{n\}}(t)$ which leads to the constitutive relation:
\begin{equation}
\frac{\Theta_{max}^{\{n\}}(t)}{M_{pad}^{\{n\}} (t)} = \frac{h}{G_z I_\theta} \frac{\tan (\lambda'')}{\lambda''}, \label{thetaMi}
\end{equation}
with $\lambda''=\omega_n / \sqrt{G_z/(\rho h^2)}$ (again non-resonant condition).
Computing the torsional second moment of area $I_\theta$ of non-circular cross sections is a difficult issue \cite{cleland}. For a rectangular beam$^($\footnote{This expression is symmetric with respect to $w \leftrightarrow e$, and for the square beam reduces to $2.25 \, e^4$.}$^)$ of width $w$ and thickness $e$, it can be written \cite{wolfbars}:
%\vspace*{-3mm}
\begin{displaymath}
I_\theta = \frac{16}{3} w e^3 \left[1 - 192 \frac{e}{w \pi^5} \sum_{j=0}^{\infty} \frac{\tanh \left( \frac{[2 j +1] \pi w}{2 e} \right) }{\left(2 j +1 \right)^5} \right]. 
\end{displaymath}
We have $\Theta_{max}^{\{n\}} (t) = \partial f_{pad}^{\{n\}} (y=0,t)/ \partial y$ and $M_{pad}^{\{n\}} (t)=E_y I_y \, \partial^2 f_{pad}^{\{n\}} (y=0,t)/ \partial y^2$ for the right foot, and the corresponding equations for the left one.
We write:
\begin{eqnarray}
  \Theta_{max}^{\{n\}}(t) &  = &   \theta_{pad}^{\{n\}} \, \frac{x_n(t)}{l} , \label{eqangle}\\
      M_{pad}^{\{n\}}(t)  &  = &  \frac{E_y I_y}{l^2}   \, x_n(t)\, M^{\left\{n\right\}}_0  \left[ 1 + M^{\left\{n\right\}}_{\theta} \theta_{pad}^{\{n\}} \right] , \label{eqMoment}
\end{eqnarray}
valid for any $\theta_{pad}^{\{n\}}$. Injecting Eqs. (\ref{eqangle}) and (\ref{eqMoment}) into Eq. (\ref{thetaMi}), we obtain:
\begin{displaymath}
\theta_{pad}^{\{n\}} = \frac{\theta^{\{n\}}_a}{1-\theta^{\{n\}}_a M^{\left\{n\right\}}_{\theta}} ,
\end{displaymath}
with $\theta^{\{n\}}_a = \left( \frac{E_y I_y \, h}{G_z I_\theta \,l} \right) \frac{\tan (\lambda'')}{\lambda''} M^{\left\{n\right\}}_0$. The parameters $M^{\left\{n\right\}}_i$ (with $i=0, \theta$) are functions of $\lambda'$ and are obtained from Eq. (\ref{padshape}), for symmetric modes: 
\begin{eqnarray*}
M^{\left\{n\right\}}_0 & = & \lambda'^2\, \frac{ \cosh\left( \frac{\lambda'}{2}\right) \sin\left( \frac{\lambda'}{2}\right) - 
\cos\left( \frac{\lambda'}{2}\right) \sinh\left( \frac{\lambda'}{2}\right) }{\cosh\left( \frac{\lambda'}{2}\right) \sin\left( \frac{\lambda'}{2}\right) + 
\cos\left( \frac{\lambda'}{2}\right) \sinh\left( \frac{\lambda'}{2}\right)} , \\
M^{\left\{n\right\}}_{\theta} & =& \frac{-2 \cos\left( \frac{\lambda'}{2}\right) \cosh\left( \frac{\lambda'}{2}\right) }{\lambda' \cosh\left( \frac{\lambda'}{2}\right) \sin\left( \frac{\lambda'}{2}\right) - \lambda'
\cos\left( \frac{\lambda'}{2}\right) \sinh\left( \frac{\lambda'}{2}\right)}.
\end{eqnarray*}
The final step of the modeling implies to compute the bending force $T^{\left\{n\right\}}_{pad} (t) = -E_y I_y $ $\partial^3 f_{pad}^{\{n\}} (y=0,t)/\partial y^3$ exerted by the paddle onto one (here, the right) foot. We define:
\begin{equation}
T^{\left\{n\right\}}_{pad} (t) = \frac{E_y I_y}{l^3}   \, x_n(t)\,  T^{\left\{n\right\}}_0  \left[ 1 + T^{\left\{n\right\}}_{\theta} \theta_{pad}^{\{n\}}   \right]. \label{eqforceinertia}
\end{equation}
%which will be used for the computation of the feet's modal shape $\Psi_{foot}^{\{n\}}(z)$, and modal parameter $\lambda^{\{n\}}$. 
For symmetric modes, Eq. (\ref{padshape}) brings, for any $\theta_{pad}^{\{n\}}$:
%\vspace*{-4mm}
\begin{eqnarray*}
T^{\left\{n\right\}}_0 & = & \lambda'^4 \,\frac{2 \sin\left( \frac{\lambda'}{2}\right)\sinh\left( \frac{\lambda'}{2}\right) }{ \lambda' \left[ \cosh\left( \frac{\lambda'}{2}\right)\sin\left( \frac{\lambda'}{2}\right) + \cos\left( \frac{\lambda'}{2}\right)\sinh\left( \frac{\lambda'}{2}\right) \right]} , \\
T^{\left\{n\right\}}_{\theta} & =& \frac{\coth\left( \frac{\lambda'}{2}\right)-\cot\left( \frac{\lambda'}{2}\right)}{2 \lambda'}.
\end{eqnarray*}
The prefactor in the flexural force $T^{\left\{n\right\}}_{pad} (t)$ is $\frac{E_y I_y}{l^3} \, x_n(t)\, \lambda'^4 =+ m_{pad} \, \omega_n^2 \, x_n(t)$, recast in $ -m_{pad} \, \ddot{x}_n(t)$: it corresponds to an {\it effective inertial} effect, with $m_{pad}=\rho (l w' e)$ the paddle mass$^($\footnote{For anti-symmetric modes, $T^{\left\{n\right\}}_{pad} (t)$ contains {\it both an inertial and an elastic} term: $\lambda'^4$ cannot be factorized in the $T^{\left\{n\right\}}_0$ expression.}$^)$. 
The same conclusion applies to the other foot as well. From Eq. (\ref{eqforceinertia}) we shall define:
\begin{equation}
 m^{\{n\}}_l = m_{pad} \, \frac{T^{\left\{n\right\}}_0}{\lambda'^4}  \left[1 + T^{\left\{n\right\}}_{\theta} \theta_{pad}^{\{n\}} \right] ,
\end{equation}
the effective mass load experienced by each foot in a symmetric flexure. $m^{\{n\}}_l$ is a function of $\lambda'$, or equivalently $\lambda^{\{n\}}$, Eq. (\ref{eqnlambda}).
%; it can be written in the form of an infinite series with respect to $\lambda'^4$, or equivalently $(\lambda^{\{n\}})^4$.
Expanding the modeling presented in Ref. \cite{JLTP_2008}, the mode shape of one foot $\Psi_{foot}^{\{n\}}(z)$ is obtained from the Euler-Bernoulli equation and the cantilever's boundary conditions, as a function of $\lambda^{\{n\}}$. The last boundary condition writes:
\begin{equation}
-E_z I_z \frac{d^3 \Psi_{foot}^{\{n\}}}{d z^3}(z=h) =   m^{\{n\}}_l   \, \omega_n^2 \, \Psi_{foot}^{\{n\}}(z=h)
\end{equation}
which finally defines the mode parameter $\lambda^{\{n\}}$ for all (here odd) $n$ (symmetric modes).

\begin{figure}
\begin{center}
\includegraphics[width=1.\linewidth,keepaspectratio]{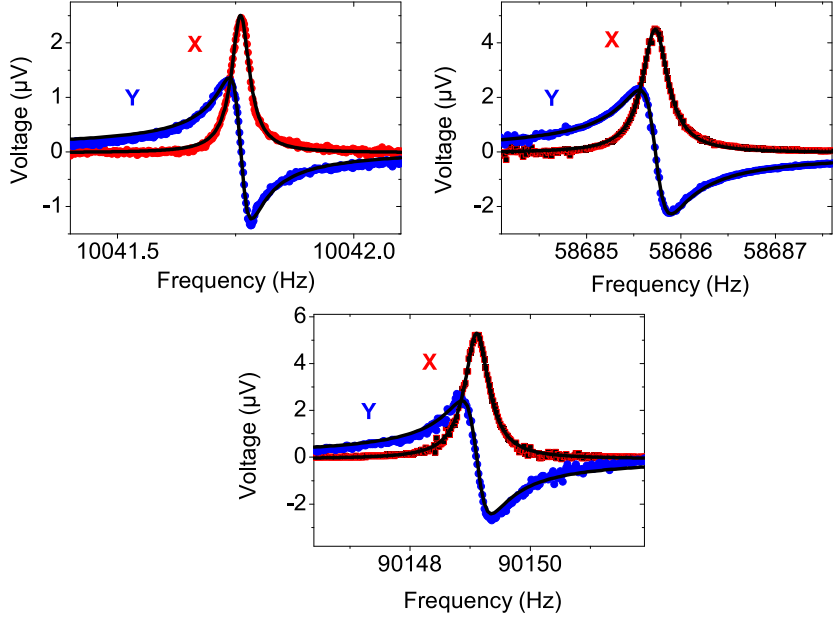}
\end{center}
\caption{(Color online) Measured resonance lines of the first symmetric out-of-plane modes, in the linear regime (4.2$~$K in vacuum). $X$ and $Y$ correspond to in-phase and quadrature components (homodyne detection, performed with a lock-in detector). From top left clockwise: first mode (field 101$~$mT and current 40$~$nA), third mode (1$~$T and 150$~$nA), and fifth mode (1$~$T and 4$~\mu$A). Black lines are Lorentzian fits.}
\label{fig2}
\end{figure}

For the first $n=1$ mode (and {\it only} for this mode), the lowest order in $\lambda'$ is enough to obtain a good accuracy, leading to $m^{\{1\}}_l \approx m_{pad}/2   + o(\lambda'^4) $.
The dynamics is essentially the one of two cantilevers loaded each by half the paddle mass, moving together because of the rigid paddle bar$^($\footnote{This intuitive result is also obtained via a simple analysis based on Rayleigh's method \cite{JLTP_2008}.}$^)$ $\Psi_{pad}^{\{1\}}(y) \approx 1 + o(\lambda'^4)$. 
For our devices, the discrepancy between $m^{\{1\}}_{l}$, $\Psi_{pad}^{\{1\}}(y)$ and their leading orders is at most a few $\%$.
At the same time, the foot torsion $\theta_{pad}^{\{1\}}$ is of the order of a few degrees only.

In Fig. \ref{fig2} we show the resonance lines measured in vacuum at 4.2$~$K using the magnetomotive scheme \cite{JLTP_2008}. The drive has been kept in the linear regime, with currents below 1$~\mu$A and fields below (or equal to) $1~$T. 
The amplitudes of the magnetomotive force and signal can be computed from the paddle shape, Eq. (\ref{padshape}).
The experimentally defined resonance frequencies are well reproduced by our modelings (within typically $\pm 10~\%$). Moreover, the quality factors of all the measured resonances are of the same order, about a fraction of a million.

\section{Conclusions}

We have characterized experimentally the 3 firsts out-of-plane symmetric flexural modes of a goalpost micro-mechanical device. The frequencies and mode shapes have been reproduced by numerical simulations and an analytic modeling without free parameters. The analytic expressions enable to compute any mode-dependent parameters needed by the experimentalist, and also prove to be extremely useful to identify the limitations imposed by all physical parameters. The linear properties are very good, with quality factors $Q$ of about $0.2 \times 10^6$ for all studied modes. 
This makes the higher flexural modes of goalpost structures perfectly suitable for experiments in quantum fluids.
The nonlinear properties of these modes shall be presented elsewhere.

\begin{acknowledgements}
We would like to thank B. Fernandez, T. Fournier, C. Blanc and O. Bourgeois for help with the fabrication of the devices, and M. N\~unez-Regueiro for financially supporting the ANSYS project via ANR grant TetraFer ANR-09-BLAN-0211. We acknowledge the support from MICROKELVIN, the EU FRP7 low temperature infrastructure grant 228464, and of the 2010 ANR French grant QNM n$^\circ$ 0404 01. 
\end{acknowledgements}

\end{document}